\documentclass[sigconf]{acmart}

\usepackage{tikz}
\usetikzlibrary{arrows.meta, positioning, shapes.misc}
\newcommand{\icon}[1]{%
  \raisebox{-0.2ex}{\includegraphics[height=1.2em]{fig/icons/icon#1.png}}%
}
\usepackage{framed}
\usepackage{xcolor}
\definecolor{framecolor}{rgb}{0.8,0.2,0.2} 
 {\endMakeFramed}

\AtBeginDocument{%
  }

\setcopyright{acmlicensed}
\copyrightyear{2018}
\acmYear{2018}
\acmDOI{XXXXXXX.XXXXXXX}

\copyrightyear{2026}
\acmYear{2026}
\setcopyright{cc}
\setcctype{by-nc-nd}
\acmConference[CHI EA '26]{Extended Abstracts of the 2026 CHI Conference on Human Factors in Computing Systems}{April 13--17, 2026}{Barcelona, Spain}
\acmBooktitle{Extended Abstracts of the 2026 CHI Conference on Human Factors in Computing Systems (CHI EA '26), April 13--17, 2026, Barcelona, Spain}
\acmDOI{10.1145/3772363.3798371}
\acmISBN{979-8-4007-2281-3/2026/04}

\begin{document}

\title[MagHeart: Exploring Playful Avatar Co-Creation and Shared Heartbeats for Icebreaking in Hybrid Meetings]{MagHeart: Exploring Playful Avatar Co-Creation and Shared Heartbeats for Icebreaking in Hybrid Meetings}

\author{Black Sun}
\authornote{Both authors contributed equally to this research.}
\affiliation{%
 \institution{Department of Computer Science\\Aarhus University}
 \city{Aarhus}
 \country{Denmark}
 }
\email{202403892@post.au.dk}

\author{Haiyang Xu}
\authornotemark[1]
\affiliation{%
 \institution{Department of Computer Science\\Aarhus University}
 \city{Aarhus}
 \country{Denmark}
 }
\email{202402604@post.au.dk}

\author{Ge Kacy Fu}
\affiliation{%
 \institution{Department of Computer Science\\Aarhus University}
 \city{Aarhus}
 \country{Denmark}
 }
\email{kacyfu@163.com}

\author{Liyue Da}
\affiliation{%
 \institution{Department of Computer Science\\Aarhus University}
 \city{Aarhus}
 \country{Denmark}
 }
\email{liyueda@cs.au.dk}

\author{Eve Hoggan}
\affiliation{%
 \institution{Department of Computer Science\\Aarhus University}
 \city{Aarhus}
 \country{Denmark}
 }
\email{eve.hoggan@cs.au.dk}


\renewcommand{\shortauthors}{B. Sun et al.}


\begin{abstract}

\end{abstract}

\begin{abstract}
Hybrid meetings often begin with social awkwardness and asymmetric participation, particularly for remote attendees who lack access to informal, co-present interaction. We present MagHeart, a multimodal system that explores symmetric icebreaking in hybrid meetings through playful LEGO-based avatar co-creation and a tangible magnetic device that represents a remote participant's heartbeat as an ambient presence cue. By combining creative co-creation with abstract bio-feedback, MagHeart rethinks how remote participants can become materially and perceptually present during meeting openings. We report findings from a scenario-based exploratory study combining quantitative and qualitative data, examining participants’ anticipated engagement, perceived social presence, and future-use intentions from both co-located and remote perspectives. Our results highlight opportunities for playful, embodied icebreakers to support early hybrid interaction, while also surfacing tensions around privacy, distraction, and contextual appropriateness. This work contributes design insights and open questions for future hybrid meeting tools that balance playfulness, embodiment, and social sensitivity.
\end{abstract}




\maketitle

\section{Introduction}

Hybrid meetings, in which some participants are co-located while others join remotely, have become a common mode of collaboration~\cite{constantinides2022future}. Despite advances in video conferencing technology, the opening moments of hybrid meetings often remain awkward and uneven~\cite{mu2024whispering, neumayr2021hybrid}. Informal greetings and small talk tend to emerge naturally among those physically present, whereas remote participants frequently struggle to find appropriate moments to join the conversation~\cite{bleakley2022bridging}. This imbalance at the beginning of meetings can shape participation dynamics for the rest of the session.

Icebreaking activities are commonly used to reduce social tension and support early interaction~\cite{kilanowski2012breaking}. However, many existing icebreakers assume physical co-presence or rely solely on verbal interaction, making them difficult to adapt to hybrid settings~\cite{csat2022web}. As a result, remote participants are often positioned as passive responders rather than active contributors during meeting openings. Designing icebreakers that are playful, low-pressure, and accessible across physical and remote roles remains an open challenge~\cite{saatcci2019hybrid, jiang2025playful}. At the same time, prior research suggests that sharing physiological signals, such as heart rate, can support social awareness and emotional understanding in mediated interaction~\cite{hassib2017heartchat,liu2017supporting, moge2022shared}. Yet exposing such signals also introduces tensions around interpretation, distraction, and privacy, particularly in professional contexts~\cite{wikstrom2021heart}. How these opportunities and tensions play out during the sensitive opening phase of hybrid meetings remains rarely understood.


In this work, we present \textit{MagHeart}, a multimodal system that explores symmetric icebreaking in hybrid meetings through playful LEGO-based avatar co-creation and a tangible magnetic device that renders a remote participant’s heartbeat as an ambient presence cue. Our contributions are: (1) the design and implementation of the \textit{MagHeart} prototype and accompanying mobile app; and (2) findings from a scenario-based exploratory study offering quantitative and qualitative insights into participants’ anticipated engagement, perceived social presence, and related design tensions in professional contexts.

\begin{figure*}[h!]
\centering
\includegraphics[width=\linewidth]{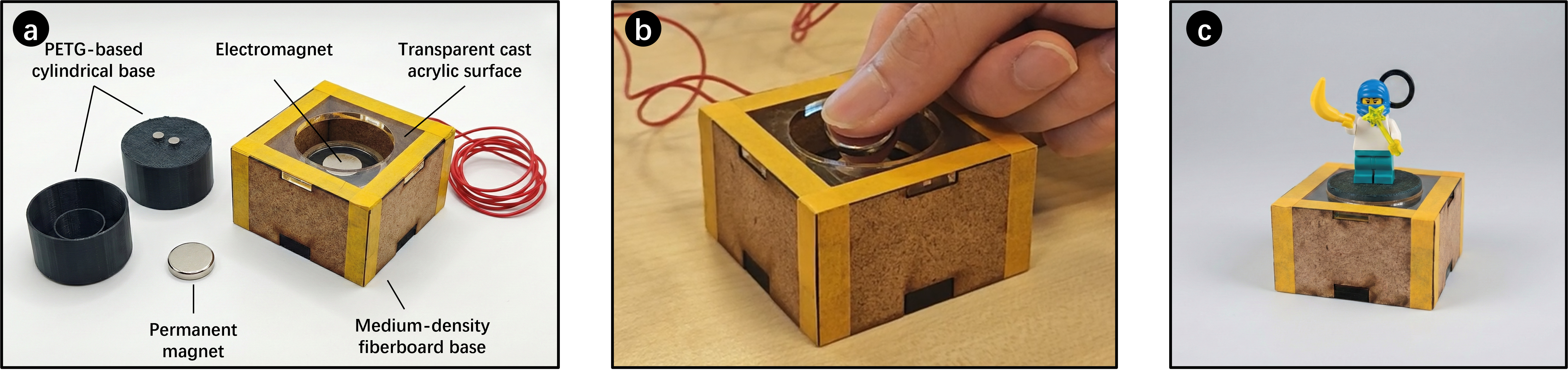}
\caption{MagHeart physical display.
(a) Exploded view of the device components, including the MDF base enclosure with an embedded electromagnet, a transparent cast acrylic top surface with a circular aperture, and a PETG-based cylindrical top object containing a permanent magnet. (b) A co-located user holds the permanent magnet to feel the remote user's pulse. (c) Assembled configuration with the LEGO figurine placed on the base.}
\label{fig:device}
\end{figure*}

\section{MagHeart}


\subsection{Design Goals}
MagHeart was designed to support inclusive and low-pressure icebreaking in hybrid meetings by combining playful avatar co-creation with physiological cues. Based on challenges identified in prior work on hybrid meetings, we defined the following design goals: (1) \textbf{DG1: Lower the barrier to early participation.} The system aims to reduce initial social awkwardness in hybrid meetings by providing low-pressure, non-verbal entry points into interaction, particularly for remote participants. (2) \textbf{DG2: Support remote participants’ agency and social presence.} Through playful avatar co-creation and the physical presence of a LEGO figure in the co-located space, the design seeks to make remote participants more visibly and materially present during icebreaking and following the hybrid meeting. (3) \textbf{DG3: Explore shared physiological cues as ambient support.} By presenting heart rate as a rhythmic signal, MagHeart explores how shared physiological cues might ease social tension during the hybrid meetings, especially for remote participants.

\subsection{System Implementation}

As shown in \autoref{fig:device}a, MagHeart combines a modular physical display with software. The physical device consists of a cubic MDF base housing an electromagnet and control electronics, and a LEGO figurine mounted on a PETG-printed base with an embedded permanent magnet. When placed above a circular aperture on the transparent acrylic surface, the figurine magnetically couples with the base. An ESP32 microcontroller modulates the electromagnet via PWM (0--255), producing subtle lifting and resistive motions that simulate a heartbeat-like rhythm and function as an ambient, tangible presence cue. The system is powered by a 12\,V supply for stable operation. The software backend is implemented in Python using FastAPI\footnote{https://fastapi.tiangolo.com}, supporting both the MagHeart mobile app and the Icebreaking Interface. Heart-rate data are captured on Apple Watch via HealthKit and transmitted through a paired iOS device to the server, which maintains the latest state and distributes updates to the physical device via server-sent events (SSE) and to interfaces via WebSocket connections.

\begin{figure*}[h!]
\centering
\includegraphics[width=\linewidth]{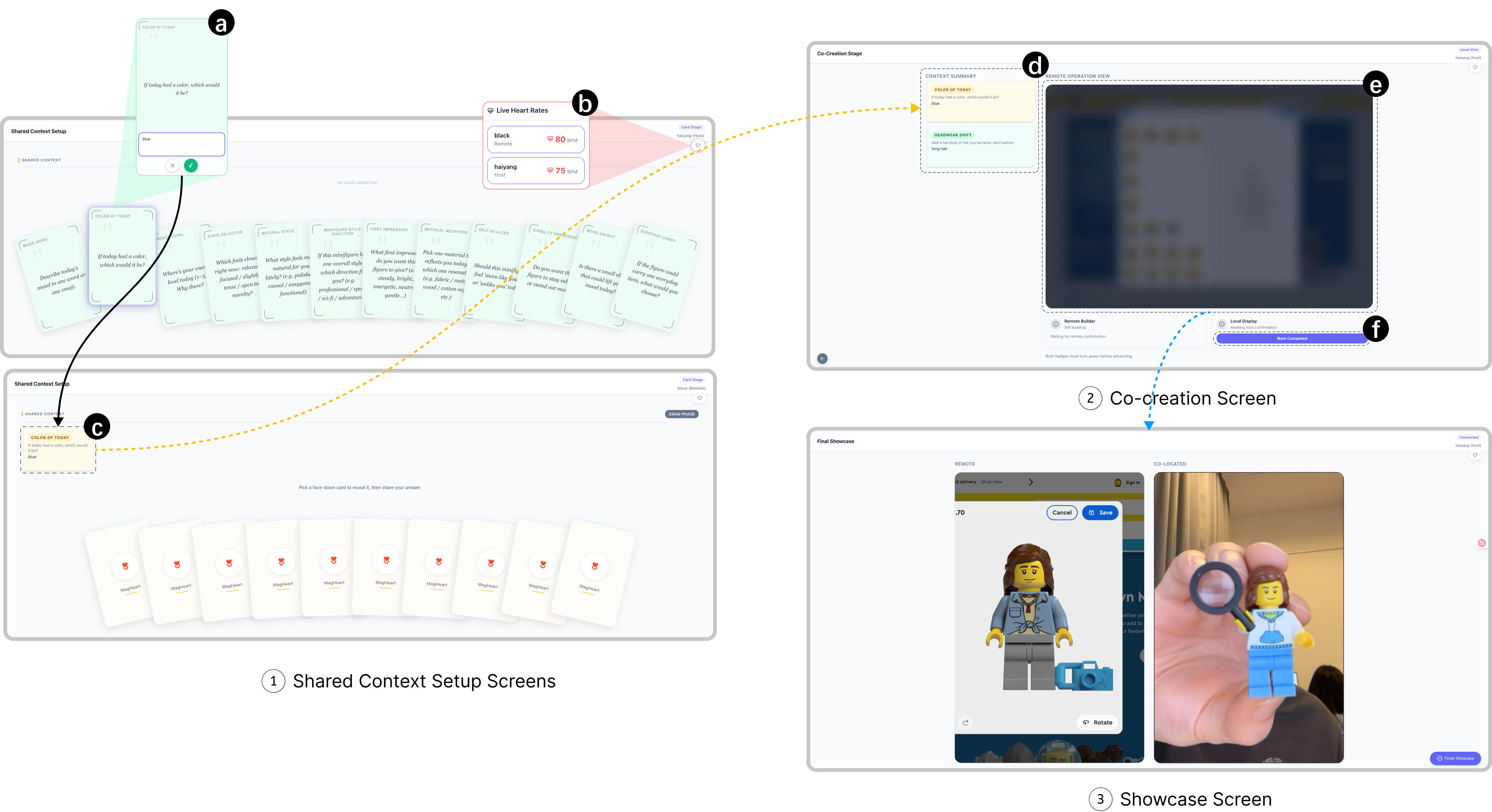}
\caption{Main screens of the MagHeart App. In the Shared Context Setup Screens~\icon{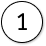}, the co-located participant selects a first-round prompt card, which is shown to the remote participant for response(~\icon{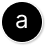} upper). A persistent heart-rate indicator~(b) remains visible throughout the interaction by clicking the ``\icon{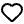}'' button. Each response is recorded as a preference tag~(c), forming part of the shared context. The second-round random cards operate in the same way(~\icon{a} bottom). In the Co-Creation Screen~\icon{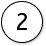}, all collected tags appear as shared context~\icon{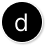}. The remote participant shares their building process through a blurred preview~\icon{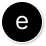}, which lets the co-located side sense ongoing activity without revealing detail. Both sides indicate progress using the complete button~\icon{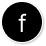}. When finished. In the Showcase Screen~\icon{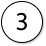}, the system presents the remote participant’s digital figure alongside the co-located group’s physical build, enabling comparison and discussion before finalizing the shared design.}
\label{fig:app}
\end{figure*}

\subsection{Interaction Flow}\label{sec:flow}

The icebreaking interaction supported by MagHeart unfolds through three stages that guide participants from initial connection to sustained presence across hybrid spaces: (1) \textbf{Establishing Connection through Tactile Heartbeat Synchronization.} At the start of the session, both co-located and remote participants wear Apple Watches to enable real-time heart-rate sensing. The co-located participant can hold a permanent magnet near the MagHeart base to physically sense the remote participant’s heartbeat through magnetic resistance and rhythmic motion (\autoref{fig:device}b). This tactile interaction provides an embodied and low-verbal entry point into the shared activity, establishing an initial sense of connection across distance. (2) \textbf{Collaborative Avatar Co-Creation.} As shown in~\autoref{fig:app}, participants then engage in cross-space co-creation of a LEGO avatar using the MagHeart app. The remote participant uses the LEGO website~\footnote{\url{https://www.lego.com/en-us/minifigure-factory}} to design, while the co-located participant simultaneously assembles the physical LEGO figure. Through negotiation and coordination, both sides collaboratively construct a tangible avatar that represents the remote participant. (3) \textbf{Persistent Presence through Rhythmic Animation.} After the co-creation phase, the completed LEGO avatar is placed on the MagHeart device (\autoref{fig:device}c). Throughout the remainder of the meeting, the avatar moves rhythmically in response to the remote participant’s real-time heart rate.

\section{Scenario-Based Exploratory Study}

\subsection{Study Procedure and Participants}

Due to logistical constraints in synchronizing multi-site hybrid meeting setups, we conducted a video-based exploratory study to examine the initial conceptual appeal and perceived social affordances of MagHeart. The study was administered as an online survey and focused on participants' anticipated experiences rather than actual system use.

The study followed a pre-post scenario-based survey design. First, participants reflected on their typical hybrid meeting experiences and reported, separately for co-located and remote roles, their levels of embarrassment, anxiety, participation, and speaking intent during meeting openings. Participants then watched a scenario-based demo video illustrating a hybrid meeting using MagHeart following the flow shown in Section~\ref{sec:flow}. After viewing the video, participants were asked to anticipate their experiences when using MagHeart in a hybrid meeting and answered the same set of questions for both co-located and remote roles. All pre- and post-scenario items were measured using 7-point Likert scales (1 = \textit{Strongly disagree}, 7 = \textit{Strongly agree}).

To capture perceived social presence associated with different design elements of MagHeart, we included items assessing anticipated social presence in relation to: (1) sensing a remote participant’s heartbeat through magnetic interaction, (2) collaboratively co-creating a LEGO avatar via the MagHeart app, and (3) observing the animated LEGO avatar in the co-located space during the meeting. Corresponding items were presented from both co-located and remote perspectives. In addition, participants reported their anticipated willingness to use MagHeart as an icebreaking tool in future hybrid meetings from both roles. We also collected qualitative feedback based on the video study.

We recruited 34 participants (P1-P34, 17 male, 15 female, 2 non-binary) through an online survey with a demo video illustrating the MagHeart concept and workflow. Participants were aged from 19 to 35 years (M = 24.18, SD = 3.13). All participants reported prior experience with hybrid meetings, with 15 indicating frequent participation.

\begin{figure*}[h!]
\centering
\includegraphics[width=\linewidth]{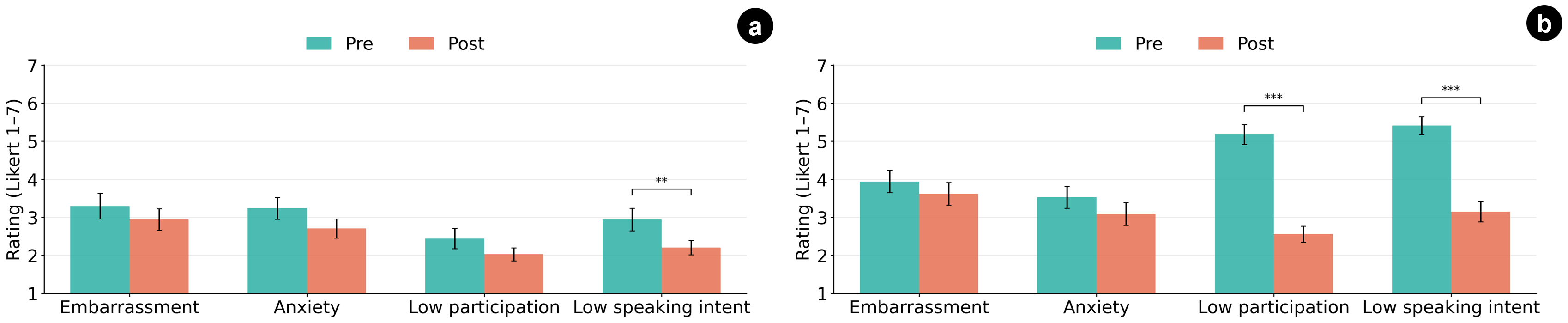}
\caption{Pre–post comparisons of participants’ anticipated experiences in hybrid meetings before (Pre) and after the MagHeart scenario (Post), measured on 7-point Likert scales (1 = strongly disagree, 7 = strongly agree; error bars indicate standard errors).
(a) Co-located perspective. (b) Remote perspective.
Asterisks indicate statistical significance (** p < .01, *** p < .001).}
\label{fig:t-test}
\end{figure*}

\begin{figure*}[h!]
\centering
\includegraphics[width=\linewidth]{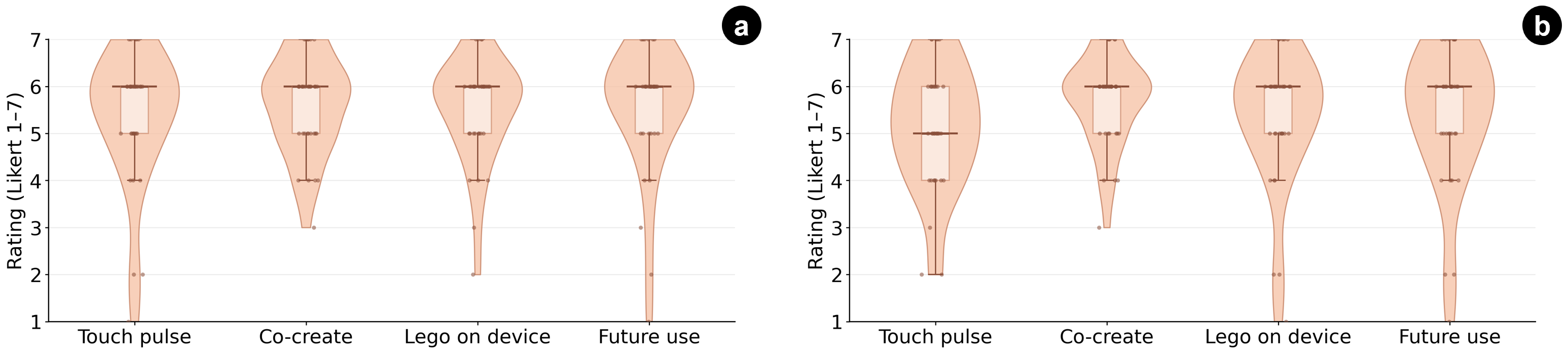}
\caption{Distribution of participants’ post-scenario ratings for perceived social presence and future-use intention after the MagHeart condition. (a) Co-located perspective. (b) Remote perspective.
Violin plots show rating distributions with overlaid medians and interquartile ranges.}
\label{fig:violin}
\end{figure*}

\subsection{Quantitative Results}

We conducted paired-samples \textit{t}-tests (two-tailed; $df = 33$; $n = 34$) to compare participants’ anticipated experiences in typical hybrid meetings without MagHeart (pre) and their anticipated experiences after MagHeart was introduced (post). Analyses were conducted separately for co-located and remote roles.

For \textbf{co-located participants}~(\autoref{fig:t-test}a), MagHeart was associated with a significant reduction in anticipated low speaking intent (pre: $M = 2.94$; post: $M = 2.21$), $t(33) = -2.89$, $p = .0068$, Cohen’s $d_z = 0.50$. Anticipated changes in embarrassment, anxiety, and low participation were not statistically reliable (all $ps \geq .089$). For \textbf{remote participants}~(\autoref{fig:t-test}b), larger pre--post differences were observed for engagement-related measures. Anticipated low participation decreased substantially after the MagHeart scenario (pre: $M = 5.18$; post: $M = 2.56$), $t(33) = -7.87$, $p < .001$, $d_z = 1.35$. Anticipated low speaking intent was also significantly lower post-scenario (pre: $M = 5.41$; post: $M = 3.15$), $t(33) = -7.02$, $p < .001$, $d_z = 1.20$. In contrast, anticipated embarrassment and anxiety did not differ significantly before and after the scenario (all $ps \geq .198$).

After the MagHeart condition, participants reported consistently high levels of perceived social presence and future-use intention. From the \textbf{co-located perspective}~(\autoref{fig:violin}a), perceived social presence of the remote participant was high across all interaction components, including touch-based pulse sensing ($M = 5.35$, $SD = 1.43$), collaborative LEGO avatar co-creation ($M = 5.59$, $SD = 1.02$), and observing the animated avatar placed on the device during the meeting ($M = 5.50$, $SD = 1.11$). Willingness to use MagHeart in future hybrid meetings was similarly high ($M = 5.53$, $SD = 1.38$). From the \textbf{remote perspective}~(\autoref{fig:violin}b), perceived social presence was also rated highly when the co-located participant sensed their heartbeat through touch ($M = 5.12$, $SD = 1.32$), during avatar co-creation ($M = 5.68$, $SD = 0.94$), and when their LEGO avatar was placed on the device and animated by heart rate ($M = 5.32$, $SD = 1.43$). Future-use willingness was comparably high ($M = 5.35$, $SD = 1.50$). Across both perspectives, ratings clustered well above the neutral midpoint of the scale (4), with median values typically between 5 and 6.

\subsection{Qualitative Results}

Open-ended responses revealed several recurring themes regarding participants’ perceptions of MagHeart’s strengths, limitations, and design opportunities. Overall, participants engaged actively with the concept and articulated both enthusiasm for its icebreaking potential and concerns about privacy, distraction, and contextual appropriateness.

\vspace{5pt}\noindent
\textbf{Playful co-creation as a low-pressure icebreaker}
Participants frequently described MagHeart’s avatar co-creation process as an effective way to ease the social awkwardness typically associated with hybrid meeting openings. The collaborative creation of a LEGO avatar was seen as shifting early interaction away from forced small talk toward a shared task, providing a natural conversational anchor. Several participants noted that this process made it easier for remote participants to become involved early, particularly by offering a concrete focus that reduced the pressure to speak spontaneously. \textit{(P1--P3, P5, P10--P13, P18, P21, P27)}

\vspace{5pt}\noindent
\textbf{Tangible and embodied cues enhancing perceived presence}
Many participants highlighted the tangible heartbeat interaction and the physical presence of the LEGO avatar as key elements that made remote participants feel more \textit{``present''} in the shared space. From the co-located perspective, sensing a remote participant’s pulse or observing the avatar’s rhythmic movement was described as making the remote participant feel less abstract and less likely to be overlooked. From a remote perspective, the combination of shared creation and a persistent physical representation was perceived as strengthening involvement and connection compared to being represented only by a video tile or a name label. \textit{(P2, P4, P6--P9, P14--P16, P19, P22, P24, P30)}

\vspace{5pt}\noindent
\textbf{Tensions around privacy, distraction, and contextual fit}
At the same time, participants raised notable concerns about sharing physiological signals in professional settings. Heart rate was widely perceived as a sensitive and potentially ambiguous signal, with several participants expressing discomfort about exposing stress or nervousness during meetings. Others worried that the continuously moving avatar could become distracting, particularly during longer or more formal discussions. Participants also questioned whether the interaction might feel more appropriate for informal or small-group meetings than for highly structured or time-constrained sessions. These reflections point to tensions around privacy control, attention management, and the contextual appropriateness of embodied biosignal sharing in hybrid work environments. \textit{(P4, P7, P11, P15, P17, P20, P23, P25--P29, P31--P34)}



\section{Discussion, Limitations and Future Work}

The findings from our scenario-based study suggest that MagHeart’s combination of playful co-creation and embodied physiological cues may offer a promising direction for addressing asymmetries in hybrid meeting openings. Quantitative results indicate that participants anticipated reduced engagement barriers, particularly for remote roles, while qualitative feedback highlights how shared making and tangible presence cues can shift early interaction away from verbal dominance toward collaborative grounding. At the same time, participants’ reflections reveal important design tensions. While tactile heartbeat synchronization and persistent avatar motion were often perceived as strengthening presence, they also raised concerns around privacy, distraction, and interpretability. Several participants further questioned whether sharing a real-time physiological signal might feel overly creepy in formal corporate settings, where professional boundaries and impression management are salient. Together, these responses underscore that physiological signals function not only as informational cues but also as socially loaded representations that require careful framing, abstraction, and user control. Designing for professional contexts may therefore involve adjustable levels of signal abstraction and alternative non-biological ambient cues to accommodate varying comfort levels and organizational cultures.


This work has several limitations. First, the study was scenario-based and relied on participants’ anticipated experiences rather than observations of real-world use. As such, the results should be interpreted as perceptions of conceptual affordances rather than validated behavioral outcomes. Second, the interaction was demonstrated primarily in a simplified meeting scenario, which may not capture the diversity and complexity of real hybrid meetings, such as larger groups or highly time-constrained settings. Future work could explore longitudinal and in-situ deployments to examine how playful and embodied icebreakers shape hybrid meetings. Design iterations may investigate alternative or more abstract representations of physiological signals, adjustable levels of visibility, and explicit privacy controls to better accommodate user comfort.

\begin{acks}
This research was partially funded by the Independent Research Fund Denmark (grant: 10.46540/4286-00295B). We are grateful to Julia Kleinau for her inspiration for this work. We also thank all participants in our exploratory study for their time, engagement, and thoughtful feedback. Finally, we appreciate the reviewers for their constructive comments and suggestions, which helped improve this paper.
\end{acks}


\bibliographystyle{ACM-Reference-Format}
\bibliography{reference}

\end{document}